\documentclass[twocolumn,showpacs,amsmath,amssymb]{revtex4}
\usepackage{graphicx}

\newcommand{\bg}{\mbox{\boldmath $\gamma$}}
\newcommand{\bt}{\mbox{\boldmath $\theta$}}

\begin{document}

\title{BTZ Black Hole in Fisher Information Spacetime}
\author{Hiroaki Matsueda\footnote{matsueda@sendai-nct.ac.jp}}
\affiliation{
Sendai National College of Technology, Sendai 989-3128, Japan
}
\date{\today}
\begin{abstract}
We examine whether we can make a black hole in Fisher information spacetime and what kind of quantum states produce the black hole solution in terms of the anti-de Sitter spacetime/conformal field theory correspondence. Here we focus on the Ba$\tilde{\rm n}$ados-Teitelboim-Zanelli black hole. There exists a mathematical representation of entanglement spectra that define the Fisher geometry as the black hole spacetime. We find that this representation is quite similar to the entanglement spectra in a conformal field theory at finite temperature except for minor corrections, and then the inverse temperature corresponds to the position of the event horizon in the Poincare coordinate.
\end{abstract}
\maketitle

Possible application of Fisher geometry to spacetime physics has a bit long story. Since the Fisher metric is defined from a microscopic model of our target system, the resulting classical spacetime emerged from the Fisher metric may answer some questions associated with quantum gravity, efficient quantum information storage, and so on. Unfortunately, most of all known works seem to lack physical interpretation, since they are based on purely mathematical or information-geometrical viewpoints for the amount of information, not physical objects. However, I have recently written a couple of papers in which some important aspects in the anti-de Sitter spacetime/conformal field theory (AdS/CFT) correspondence in string theory can be well chaptured by the Fisher geometry~\cite{Maldacena,Matsueda,Matsueda2,Matsueda3}. Then, it was quite important to make our physical standpoint clear, and for this purpose I have used the entanglement entropy scaling in CFT. In this context, we would like to know more about functionality of the geometry by taking another famous examples in AdS/CFT.

Here we focus on the existence of the Ba$\tilde{\rm n}$ados-Teitelboim-Zanelli (BTZ) black hole in the Fisher information spacetime. The BTZ black hole geometry is the solution of the vacuum Einstein field equation in $(2+1)$ dimension with the negative cosmological constant~\cite{BTZ,BTZ2}. In view of AdS/CFT, its dual field theory is $(1+1)$-dimensional CFT, and the presence of the black hole corresponds to finite-temperature effects on CFT. As will be later discussed, the Fisher metric in the present case is defined from the entanglement spectra of a quantum field theory, and thus the spectra are determined so that the Fisher metric becomes equal to the BTZ metric. Then, we would like to ask whether the spectra are consistent with the CFT results at finite temperatures. We will find that this is actually true except for minor difference between them. Thus, we think that the consistency between this information-geometrical tool and AdS/CFT becomes more and more reliable than the previous situation.

Let us start with a quantum state $\left|\psi\right>$ defined on $(1+1)$-dimensional flat Minkowski spacetime $\mathbb{R}^{1,1}$. We devide the whole system into two spatial regions $A$ and $\bar{A}$. Then, $\left|\psi\right>$ is represented by the Schmidt decomposition or the singular value decomposition (SVD) as
\begin{eqnarray}
\left|\psi\right>=\sum_{n}\lambda_{n}(\xi,t,x)\left|n\right>_{A}\otimes\left|n\right>_{\bar{A}},
\label{wf}
\end{eqnarray}
where $\{\left|n\right>_{A}\}$ and $\{\left|n\right>_{\bar{A}}\}$ are the Schmidt bases for two subsystems $A$ and $\bar{A}$, respectively. The Schmidt coefficient or the SVD spectrum $\lambda_{n}$ is a function of correlation length $\xi$ of $\left|\psi\right>$, time $t$, and the boundary position $x$ between $A$ and $\bar{A}$. These parameters are labeled as $\bt=\left(\xi,t,x\right)$. It is well-known in terms of density matrix renormalization group (DMRG) that the index $n$ distinguishes different length-scale physics~\cite{White}. Later we will see that the presence of $\xi$ is crucial for the emergence of the radial axis of AdS. We normalize the Schmidt coefficient so that $\left|\psi\right>$ is normalized as
\begin{eqnarray}
\langle\psi|\psi\rangle=\sum_{n}\left|\lambda_{n}(\bt)\right|^{2}=1. \label{conservation}
\end{eqnarray}

The dual gravity theory is constructed by the Fisher metric defined from the entanglement entropy of the original quantum state. The entropy is defined by
\begin{eqnarray}
S(\bt)=-\sum_{n}\left|\lambda_{n}(\bt)\right|^{2}\log\left|\lambda_{n}(\bt)\right|^{2},
\end{eqnarray}
where the entanglement entropy $S(\bt)$ is a function of $\bt$ due to the $\bt$ dependence on the singular value spectrum $\lambda_{n}(\bt)$. We also define the entanglement spectrum $\gamma_{n}(\bt)$ and the expectation value of a quantity $O_{n}(\bt)$ as
\begin{eqnarray}
\gamma_{n} &=& -\log\left|\lambda_{n}(\bt)\right|^{2}, \\
\left<\mbox{\boldmath $O$}\right> &=& \sum_{n}\left|\lambda_{n}(\bt)\right|^{2}O_{n}(\bt),
\end{eqnarray}
and then the entropy is represented as
\begin{eqnarray}
S=\left<\bg\right>.
\end{eqnarray}

In order to introduce the Fisher metric, we take the second derivative of $S$ by $\bt$. We abbreviate $\partial/\partial\theta^{\mu}$ as $\partial_{\mu}$. The first derivative of $S$ is given by
\begin{eqnarray}
-\partial_{\nu}S &=& \sum_{n}\left(\left(\partial_{\nu}\left|\lambda_{n}\right|^{2}\right)\log\left|\lambda_{n}\right|^{2}+ \partial_{\nu}\left|\lambda_{n}\right|^{2}\right) \nonumber \\
&=& \sum_{n}\left(\partial_{\nu}\left|\lambda_{n}\right|^{2}\right)\log\left|\lambda_{n}\right|^{2},
\end{eqnarray}
where the second term in the right hand side vanishes, since $\sum_{n}\partial_{\nu}\left|\lambda_{n}\right|^{2}=\partial_{\nu}1=0$. The second derivative is also given by
\begin{eqnarray}
-\partial_{\mu}\partial_{\nu}S &=& \sum_{n}\frac{1}{\left|\lambda_{n}\right|^{2}}\left(\partial_{\mu}\left|\lambda_{n}\right|^{2}\right)\left(\partial_{\nu}\left|\lambda_{n}\right|^{2}\right) \nonumber \\
&& + \sum_{n}\left(\partial_{\mu}\partial_{\nu}\left|\lambda_{n}\right|^{2}\right)\log\left|\lambda_{n}\right|^{2} \nonumber \\
&=& \sum_{n}\left|\lambda_{n}\right|^{2}\left(\partial_{\mu}\gamma_{n}\right)\left(\partial_{\nu}\gamma_{n}\right) \nonumber \\
&& + \sum_{n}\left|\lambda_{n}\right|^{2}\left(\partial_{\mu}\partial_{\nu}\gamma_{n}-\left(\partial_{\mu}\gamma_{n}\right)\left(\partial_{\nu}\gamma_{n}\right)\right)\gamma_{n} \nonumber \\
&=& g_{\mu\nu}+h_{\mu\nu},
\end{eqnarray}
where $g_{\mu\nu}$ and $h_{\mu\nu}$ are respectively defined by
\begin{eqnarray}
g_{\mu\nu} &=& \left<\partial_{\mu}\bg\partial_{\nu}\bg\right>, \label{mnS} \\
h_{\mu\nu} &=& \left<\bg\left(\partial_{\mu}\partial_{\nu}\bg-\partial_{\mu}\bg\partial_{\nu}\bg\right)\right>.
\end{eqnarray}
Starting with the relation $\left<\partial_{\nu}\bg\right>=0$, we find
\begin{eqnarray}
g_{\mu\nu} &=& \left<\partial_{\mu}\bg\partial_{\nu}\bg\right> = \left<\partial_{\mu}\partial_{\nu}\bg\right>.
\end{eqnarray}
The second term $h_{\mu\nu}$ is thus vanishing in the strong coupling limit in which all of the singular values play an equal role on this term. Furthermore it is clear in mean-field decoupling that $h_{\mu\nu}\simeq\left<\bg\right>\left<\partial_{\mu}\partial_{\nu}\bg-\partial_{\mu}\bg\partial_{\nu}\bg\right>=0$. Therefore, $g_{\mu\nu}$ is crucial for the metric, and we know
\begin{eqnarray}
g_{\mu\nu}=\left<\partial_{\mu}\bg\partial_{\nu}\bg\right>\simeq -\partial_{\mu}\partial_{\nu}S(\bt).
\end{eqnarray}
The right hand side of Eq.~(\ref{mnS}) is nothing but the Fisher metric in terms of information geometry~\cite{Amari}. Actually, we calculate the Kullback-Leibler measure
\begin{eqnarray}
D(\bt) &=& \sum_{n}\left|\lambda_{n}(\bt)\right|^{2}\left(\gamma_{n}(\bt)-\gamma_{n}(\bt+d\bt)\right) \nonumber \\
&=& \frac{1}{2}\left<\partial_{\mu}\bg\partial_{\nu}\bg\right>d\theta^{\mu}d\theta^{\nu}, \label{KL}
\end{eqnarray}
and thus we know that $g_{\mu\nu}$ is an appropriate metric tensor.

Let us move to black hole configuration. The BTZ black hole is represented by the following metric
\begin{eqnarray}
ds^{2}=\frac{1}{z^{2}}\left(-f(z)dt^{2}+\frac{dz^{2}}{f(z)}+dx^{2}\right).
\end{eqnarray}
Hereafter we change the notation as $\xi\rightarrow z$ in accordance with standard notation. The factor $f(z)$ is defined by
\begin{eqnarray}
f(z)=1-\left(\frac{z}{z_{0}}\right)^{2},
\end{eqnarray}
and the position $z=z_{0}$ denotes the event horizon. First we consider a constant-$t$ surface, since the effect of the entanglement dynamics on the scaling formula of the entanglement entropy is just additive to the equillibrium result.

It is convenient to take a complex coordinate as
\begin{eqnarray}
w &=& x+ig(z) , \\
\bar{w} &=& x-ig(z) ,
\end{eqnarray}
with a $z$-dependent real function $g(z)$. We recognize
\begin{eqnarray}
dwd\bar{w} = dx^{2}+\left(\frac{dg(z)}{dz}\right)^{2}dz^{2},
\end{eqnarray}
and we take
\begin{eqnarray}
\frac{dg(z)}{dz}=\frac{1}{\sqrt{f(z)}}.
\end{eqnarray}
Integrating this differential equation, we obtain
\begin{eqnarray}
g(z)=\int\frac{dz}{\displaystyle \sqrt{1-\left(\frac{z}{z_{0}}\right)^{2}}}=z_{0}\sin^{-1}\left(\frac{z}{z_{0}}\right).
\end{eqnarray}
In the complex coordinate~\cite{Matsueda}, the Fisher metric is represented as
\begin{eqnarray}
ds^{2}&=&\left<\partial_{w}\bg\partial_{w}\bg\right>dw^{2} + \left<\partial_{\bar{w}}\bg\partial_{\bar{w}}\bg\right>d\bar{w}^{2} \nonumber \\
&&+2\left<\partial_{w}\bg\partial_{\bar{w}}\bg\right>dwd\bar{w}.
\end{eqnarray}
The entanglement spectrum in the complex coordinate is described dy the sum of holomorphic and anti-holomorphic parts. The first derivative of the entanglement spectrum, $\partial_{w}\gamma$, is expanded to the Laurent modes,
\begin{eqnarray}
\partial_{w}\gamma_{n}(w,\bar{w}) &=& \sum_{l\in\mathbb{Z}}h_{nl}w^{-l-1} , \\
\partial_{\bar{w}}\gamma_{n}(w,\bar{w}) &=& \sum_{l\in\mathbb{Z}}\bar{h}_{nl}\bar{w}^{-l-1} ,
\end{eqnarray}
and the leading order of the Laurent expansion of $\gamma$ is given by
\begin{eqnarray}
\gamma_{n} \simeq g_{n} + h_{n}\log w + \bar{h}_{n}\log\bar{w},
\end{eqnarray}
with real and complex constants, $g_{n}$ and $h_{n}$, respectively. These logarithmic terms directly correspond to the entanglement entropy in the CFT back ground, since the expectation value of the entanglement spectrum is equal to the entanglement entropy. We assume
\begin{eqnarray}
h_{n}=\alpha_{n}+i\beta_{n} \;\; , \;\; \left<\alpha^{2}\right>=\left<\beta^{2}\right> \;\; , \;\; \left<\alpha\beta\right>=0.
\end{eqnarray}
Then, the metric is given by
\begin{eqnarray}
ds^{2} &=& \frac{\left<h^{2}\right>}{w^{2}}dw^{2} + \frac{\left<\bar{h}^{2}\right>}{\bar{w}^{2}}d\bar{w}^{2} + 2\frac{\left<h\bar{h}\right>}{|w|^{2}}dwd\bar{w} \nonumber \\
&=& \frac{4\left<\alpha^{2}\right>}{|w|^{2}}\left(\frac{dz^{2}}{f(z)}+dx^{2}\right).
\end{eqnarray}
Here we have
\begin{eqnarray}
\left|w\right|^{2} = x^{2}+g^{2}(z) \sim z^{2},
\end{eqnarray}
for $x<z<z_{0}$. Then, we find that the entanglement entropy is given by
\begin{eqnarray}
S=\left<\bg\right>\simeq \Delta\log g(z)=\Delta\log\left(\frac{z_{0}}{\epsilon}\sin^{-1}\left(\frac{z}{z_{0}}\right)\right),
\end{eqnarray}
where $\epsilon$ is a UV cut-off, and we take the conformal dimension as
\begin{eqnarray}
\Delta=\left<h\right>+\left<\bar{h}\right>.
\end{eqnarray}

It should be noted that the result looks different from the CFT one at finite temperature. In order to relate this result with standard CFT one, once we expand $\sin^{-1}y$  into a power series. Then, we find
\begin{eqnarray}
\sin^{-1}y = y+\frac{1}{6}y^{3}+\cdots \simeq \sinh y.
\end{eqnarray}
Finally, the entanglement entropy is represented as
\begin{eqnarray}
S \simeq \Delta\log\left(\frac{z_{0}}{\epsilon}\sinh\left(\frac{z}{z_{0}}\right)\right).
\end{eqnarray}
This is consistent with the CFT result~\cite{Calabrese}, if we identify $z=\xi$ as the inverse temperature $\beta/2\pi$ and $\Delta=c/3$ with the central charge $c$. We may need to discuss more about why $\sin^{-1}(z/z_{0})$ appears instead of $\sinh(z/z_{0})$. In the present stage, it is unclear whether the Fisher geometry has somehow different properties from the standard spacetime physics or whether this difference is just coming from approximation and other minor corrections.

We can also take account of entanglement dynamics by our representation. We may take
\begin{eqnarray}
S(t)=S(t=0)+\kappa \frac{1}{z^{2}}f(z)t^{2},
\end{eqnarray}
with a constant $\kappa$. This is because
\begin{eqnarray}
g_{tt}\simeq -\partial_{t}\partial_{t}S(t)\propto -\frac{1}{z^{2}}f(z).
\end{eqnarray}
In the limit of $z_{0}\rightarrow\infty$, the additional term becomes $(t/\xi)^{2}$, and this is consistent with the recent numerical result~\cite{MGN}.

Summarizing, we have examined the BTZ black hole in the Fisher information spacetime. By solving the inverse problem with the help of complex analysis, we have found that the black hole is actually corresponding to the finite-temperature CFT. The point is that the Fisher geometry can describe the CFT result almost correctly, although the Fisher geometry is just the information-geometrical tool. The present result would shed new light on recent works on gravitational dynamics emerged from entanglement entropy~\cite{Wong,Nima,Takayanagi1,Takayanagi2,Faulkner,Banerjee}. On the other side, our original result is somehow deformed from the well-known CFT result, and thus we must take more sophisticated approach to resolve this difference as a future work.

\end{document}